\documentclass[journal,twocolumn]{IEEEtran}
\usepackage{}
\usepackage{graphicx}
\usepackage{epstopdf}
\usepackage{amssymb}
\usepackage{amsfonts}
\usepackage{amsmath}
\usepackage{algorithm}
\usepackage{algorithmic}
\usepackage{subeqnarray}
\usepackage{cases}
\usepackage{bm}
\usepackage{color}
\usepackage{subfigure,amsmath,amssymb,cite}
\usepackage{float}
\ifCLASSINFOpdf
\else
\fi
\begin{document}
\title{Low-complexity Joint Phase Adjustment and Receive Beamforming for Directional Modulation Networks via IRS\\ }

\author{~Rongen~Dong, Shaohua~Jiang, Xinhai~Hua, Yin~Teng, Feng~Shu, and Jiangzhou Wang,\emph{ Fellow, IEEE}
\thanks{This work was supported in part by the National Natural Science Foundation of China (Nos. 62071234, 62071289, and 61972093), the Hainan Province Science and Technology Special Fund (ZDKJ2021022),  the Scientific Research Fund Project of Hainan University under Grant KYQD(ZR)-21008 and KYQD(ZR)-21007, and the National Key R\&DProgram of China under Grant 2018YFB1801102 \emph{(Corresponding author: Feng Shu)}.}
\thanks{Rongen~Dong is with the School of Information and Communication Engineering, Hainan University,~Haikou,~570228, China.}
\thanks{Shaohua~Jiang is with the school of Electrical and Mechanical Engineering, Weifang Vocational College, ~Weifang, 262737, China.}
\thanks{Xinhai~Hua is with the Corporation of Zhongxing Telecommunication Equipment, Nanjing, 210094, China.}
\thanks{Yin~Teng is with the School of Electronic and Optical Engineering, Nanjing University of Science and Technology, Nanjing, 210094, China.}
\thanks{Feng Shu is with the School of Information and Communication Engineering, Hainan University, Haikou, 570228, China, and also with the School of Electronic and Optical Engineering, Nanjing University of Science and Technology, Nanjing, 210094, China (e-mail: shufeng0101@163.com).}
\thanks{Jiangzhou Wang is with the School of Engineering, University of Kent, Canterbury CT2 7NT, U.K. (e-mail: {j.z.wang}@kent.ac.uk).}
}
\maketitle

\begin{abstract}
Intelligent reflecting surface (IRS) is a revolutionary and low-cost technology for boosting the spectrum and energy efficiencies in future wireless communication  network. In order to create controllable multipath transmission in the conventional line-of-sight (LOS) wireless communication environment, an IRS-aided directional modulation (DM) network is considered. In this paper, to improve the transmission security of the system  and  maximize the receive power sum (Max-RPS), two alternately optimizing schemes of jointly designing receive beamforming (RBF) vectors and IRS phase shift matrix (PSM) are proposed: Max-RPS using general alternating optimization (Max-RPS-GAO) algorithm and Max-RPS using zero-forcing (Max-RPS-ZF) algorithm. Simulation results show that, compared with the no-IRS-assisted scheme and the no-PSM optimization scheme, the proposed IRS-assisted Max-RPS-GAO method and Max-RPS-ZF method can significantly improve the secrecy rate (SR) performance of the DM system. Moreover, compared with the Max-RPS-GAO method, the proposed Max-RPS-ZF method has a faster convergence speed and a certain lower computational complexity.
\end{abstract}

\maketitle
\section{Introduction}
The broadcast characteristic of wireless medium makes the transmission of information vulnerable to eavesdropping \cite{Mukherjee2014Principles, Yang2015Articial}.  As a complement to high-layer encryption techniques, physical layer security (PLS), which safeguards data confidentiality based on the information-theoretic approaches, has attracted wide attentions from academia and industry in the past decades
\cite{Pecchetti2019Channel, WangDistributed, ZhaoAnti, Yang2013MIMO, Zou2016Relay, Hamamreh2019Classifications,  WangSecuring, WuSecure, Hong2013Enhancing}.
The core principle of PLS is to exploit the characteristics of wireless channels to guarantee secure communication in the presence of eavesdroppers\cite{Cheng2021Physical}. Directional modulation (DM), as an advanced and promising PLS communications technique, has been regarded as a useful method for fifth generation (5G) millimeter-wave wireless communications\cite{Wang2018Hybrid, Nusenu2019Development}. DM employs signal processing technologies like beamforming and artificial noise (AN) in radio frequency frontend or baseband, so that the signal in the desired direction can be recovered as fully as possible, while the signal constellation diagram in the undesired direction is distorted\cite{Daly2010Demonstration, Shu2018Secure, Qiu2021Security}.

In \cite{Daly2009Directional}, a DM scheme using the phased arrays to generate modulation was presented, and the secure transmission was achieved since the signal was direction-dependent and purposely distorted in the undesired directions.  In \cite{Feng2016Robust}, a robust synthesis method for multi-beam DM in broadcasting systems was proposed, a robust maximum signal-to-leakage-noise ratio and maximum the signal-to-AN ratio scheme for the desired and eavesdroppers directions were presented, and an obvious bit error rate (BER) improvement over the existing orthogonal projection method along the desired direction for a given signal-to-noise ratio (SNR) was achieved.  The authors in \cite{ZhangBMulti} developed a multi-carrier based DM framework using antenna arrays, which achieved simultaneous data transmission over multiple frequencies, and a higher data rate was achieved. In \cite{Zhang2020Impact}, the impact of imperfect angle estimation on spatial and directional modulation system was investigated, with the help of the union bound and statistics theory, the average BERs for the legitimate user and eavesdropper were derived.  A scenario for DM network with a full-duplex malicious attacker was considered in \cite{Teng2022Low}, and three receive beamforming methods were proposed for enhancing the security performance. In \cite{Dong2022Performance1}, the authors investigated the performance of a hybrid analog and digital DM with mixed phase shifters. The closed-form expressions of signal-to-interference-plus-noise ratio, secrecy rate (SR), and BER were derived based on the law of large numbers.

The rapid development of wireless networks will lead to serious energy consumption. Different from the relay\cite{Multiuser2016Li, Energy2014Li, A2009Li}, intelligent reflecting surface (IRS) has been considered to be a promising green and cost-effective solution to improve the performance of wireless communication in recent years\cite{Wu2020Towards, Huang2019Reconfigurable, Di2020Smart, Tang2020MIMO, Dong2022Performance}.
The IRS can change the phase shift of the incident electromagnetic wave, thereby intelligently reconfiguring the signal propagation environment, enhancing the power of the required received signal or suppressing interference signals\cite{wuqq-zongshu}. A challenging scenario was considered in \cite{Cui2019Secure}, where the eavesdropping channel was stronger than the legitimate user channel and they were highly correlated. The access point (AP) transmit beamforming and IRS reflect beamforming were jointly designed to maximize the SR. In \cite{Wu2019Intelligent}, an IRS-aided single-cell wireless system was investigated. To minimize the total transmit power at the AP, based on semidefinite relaxation and alternating optimization methods, efficient schemes were proposed to make a tradeoff between  system performance and computational complexity. The authors in \cite{Pan2020Multicell} employed an IRS at the cell boundary to enhance the cell-edge user performance in multi-cell communication systems, in order to maximize the weighted sum rate of all users, the block coordinate descent algorithm was proposed for alternately optimizing the transmit precoding matrices at the base stations (BSs) and the passive beamforming at the IRS. In \cite{Wang2022Beamforming}, an IRS-aided decode-and-forward relay network was proposed, three high-performance beamforming methods were designed to maximize receive power. In \cite{Wang2022Intelligent}, the authors investigated an IRS aided millimeter wave (mmWave) communication system using hybrid precoding at the BS. Based on the rank-one property of mmWave channels, the closed-form solution of the approximated maximum received power of user was derived.

To overcome the limitation that only one bit stream can be transmitted between the BS and user in the conventional DM networks and create controllable multipath transmission in the line-of-sight (LOS) scenario, employing IRS in DM network has been considered. An IRS-assisted DM system was proposed in\cite{Lai2020Directional} to utilize the multipath propagation environment for enhancing the PLS, and the closed-form expression for the SR was derived. The authors in \cite{Dong2022Beamforming} considered a double-IRS-aided two-way DM network, two transmit beamforming methods were proposed to enhance the secrecy sum rate (SSR), and an effective power allocation scheme was designed to maximize the SSR performance. In \cite{ShuEnhanced2021}, in order to maximize the SR performance of IRS-assisted DM system, two alternating iterative methods, called general alternating iterative and null-space projection, were proposed. The former was of high-performance and the latter was of low-complexity.

However, the authors in \cite{ShuEnhanced2021} aimed to maximize SR of the system by designing the transmit beamforming at the transmitter and the phase shift matrix (PSM) at the IRS, without considering the receive beamforming.  Therefore, in this paper, the design of the receive beamforming is focused on and two receive beamforming (RBF) methods are proposed to improve the SR performance by taking the phase optimization of IRS into account.  The main contributions of this paper are summarized as follows:
\begin{enumerate}
\item To enhance the SR performance in the traditional DM networks, an IRS-aided DM system is considered. To improve the transmission security of the system  and reduce the detection complexity of receiver, a general alternating optimization (GAO) of  maximizing the receive power sum (Max-RPS) algorithm, called Max-RPS-GAO, is proposed firstly to attain two RBF vectors and the PSM of IRS by making use of the Rayleigh-Ritz theorem and derivative operation. Its basic idea is to alternatively optimize the IRS PSM and the two RBF vectors.
\item To receive confidential messages (CMs) from the direct path and the IRS reflected path independently, a Max-RPS using zero-forcing (Max-RPS-ZF) method is proposed. Here, the first RBF vector forces the signal directly from Alice to zero, and  the second one forces the signal from IRS to zero.  Simulation results show that, compared with the no-IRS-assisted scheme and  no-PSM optimization scheme, the proposed IRS-assisted Max-RPS-GAO algorithm and Max-RPS-ZF algorithm can improve the SR performance of the DM system. Furthermore, compared to the the Max-RPS-GAO, the proposed Max-RPS-ZF algorithm converges faster with a lower computational complexity.
%
%
\end{enumerate}

The remainder of this paper is organized as follows. Section \ref{S2} describes the IRS-based DM system model. The Max-RPS-GAO and Max-RPS-ZF methods are proposed in Section \ref{S3} and Section \ref{S4}, respectively. Section \ref{S5} presents the simulation results and related analysis. Finally, we draw conclusions in Section \ref{S6}.

\emph{Notations:} throughout this paper, scalar, vector, and matrix are denoted by letters of lower case, bold lower case, and bold upper case, respectively. Symbols $(\cdot)^T$, $(\cdot)^H$, $(\cdot)^{-1}$, $(\cdot)^{\dagger}$, and $\text{det}\{\cdot\}$ are transpose, conjugate transpose, inverse, pseudo-inverse, and matrix determinant, respectively. The notation $\textbf{I}_N$ denotes the $N\times N$ identity matrix. The sign $\mathbf{0}_{N\times M}$ represents the $N\times M$ matrix of all zeros.

\section{System Model and Problem Formulation}\label{S2}
\subsection{System Model }
\begin{figure}[htbp]
  \centering
  \includegraphics[width=0.49\textwidth]{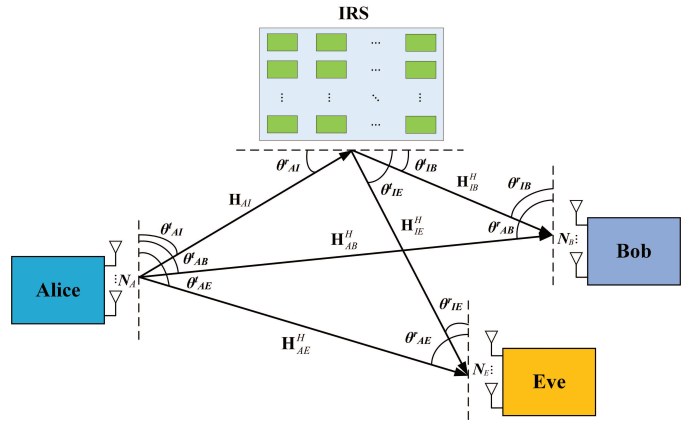}
  \caption{Block diagram for IRS-based directional modulation network.}
  \label{systemmodel}
\end{figure}
As shown in Fig.~\ref{systemmodel}, an IRS-aided DM communication network is considered in this paper, where the transmitter (Alice) is equipped with $N_A$ antennas, IRS is equipped with $M$ low-cost passive reflecting elements, legitimate user (Bob) and eavesdropper (Eve) are equipped with $N_B$ and $N_E$ antennas, respectively. In the following, it is assumed that the signals reflected more than once by the IRS are omitted due to the significant path loss \cite{Wu2019Intelligent}, and the channels from Alice to IRS, Alice to Bob, Alice to Eve, IRS to Bob, and IRS to Eve are the LOS channels.

The transmit baseband signal is
\begin{equation}\label{s}
\mathbf{s}=\sqrt{\beta_1 P_s}\mathbf{v}_{1}x_{1}+\sqrt{\beta_2 P_s}\mathbf{v}_{2}x_{2}+\sqrt{\beta_3 P_s}\mathbf{P}_{AN}\mathbf{z},
\end{equation}
where $P_s$ is the total transmit power, $\beta_1$, $\beta_2$ and $\beta_3$ denote the power allocation parameters of {CMs} and {AN}, respectively, and $\beta_1+\beta_2+\beta_3=1$. $\mathbf{v}_1\in \mathbb{C}^{N_A\times 1}$ and $\mathbf{v}_2\in \mathbb{C}^{N_A\times 1}$ represent the beamforming vector of forcing the two CMs to the desired user Bob, where $\mathbf{v}_{1}^{H}\mathbf{v}_{1}=1$ and $\mathbf{v}_{2}^{H}\mathbf{v}_{2}=1$.  $\mathbf{P}_{AN}$ denotes the projection matrix for controlling the direction of AN. $x_1$ and $x_2$ are CMs which satisfy $\mathbb{E}\left[\|x_1\|^2\right]=1$ and $\mathbb{E}\left[\|x_2\|^2\right]=1$. $\mathbf{z}$ represents the AN vector with complex Gaussian distribution, i.e., $\mathbf{z}\sim\mathcal{C}\mathcal{N}(0,~\mathbf{I}_{N_A})$.

The received signal at Bob is given by
\begin{align}\label{yb}
y_{Bi}&=\mathbf{u}^H_{Bi}\left[(\sqrt{g_{AIB}}\mathbf{H}^{H}_{IB}\boldsymbol{\Theta}\mathbf{H}_{AI}+
\sqrt{g_{AB}}\mathbf{H}^{H}_{AB})\mathbf{s}+\mathbf{n}_{B}\right] \nonumber\\
&=\mathbf{u}^H_{Bi}\big[\sqrt{\beta_1 P_s} (\sqrt{g_{AIB}}\mathbf{H}^{H}_{IB}\boldsymbol{\Theta}\mathbf{H}_{AI}+\sqrt{g_{AB}}\mathbf{H}^{H}_{AB})\mathbf{v}_{1}x_{1} \nonumber\\
&~~~+ \sqrt{\beta_2 P_s}\left(\sqrt{g_{AIB}}\mathbf{H}^{H}_{IB}\boldsymbol{\Theta}\mathbf{H}_{AI}+\sqrt{g_{AB}}\mathbf{H}^{H}_{AB} \right)\mathbf{v}_{2}x_{2} \nonumber\\
&~~~+\sqrt{\beta_3 P_s}
\left(\sqrt{g_{AIB}}\mathbf{H}^{H}_{IB}\boldsymbol{\Theta}\mathbf{H}_{AI}+\sqrt{g_{AB}}\mathbf{H}^{H}_{AB} \right)\mathbf{P}_{AN}\mathbf{z} \nonumber\\
&~~~+\mathbf{n}_{B}\big],~~i=1,2,
\end{align}
where $\mathbf{u}_{Bi}\in\mathbb{C}^{N_B\times 1}$ represents the receive beamforming vector of Bob, $\mathbf{H}^H_{IB}=\mathbf{h}(\theta^r_{IB})\mathbf{h}^H(\theta^t_{IB})\in\mathbb{C}^{N_B\times M}$ represents the IRS-to-Bob channel, $\boldsymbol{\Theta}=\text{diag}(e^{j\varphi_{1}},\cdots,e^{j\varphi_{m}},\cdots, e^{j\varphi_{M}})$ is a diagonal matrix with the phase shift $\varphi_{m}$ incurred by the $m$-th reflecting element of the IRS, $\mathbf{H}_{AI}=\mathbf{h}(\theta^r_{AI})\mathbf{h}^H(\theta^t_{AI})\in\mathbb{C}^{M\times N_A}$ represents the Alice-to-IRS channel, $\mathbf{H}^H_{AB}=\mathbf{h}(\theta^r_{AB})\mathbf{h}^H(\theta^t_{AB})\in\mathbb{C}^{N_B\times N_A}$ represents the Alice-to-Bob channel, and $\mathbf{n}_B\sim\mathcal{C}\mathcal{N}(\mathbf{0},\sigma_B^2\mathbf{I}_{N_B})$ denotes the complex additive white Gaussian noise (AWGN) at Bob.
$g_{AIB}=g_{AI} g_{IB}$ denotes the equivalent path loss coefficient of Alice-to-IRS channel and IRS-to-Bob channel, and $g_{AB}$ is the path loss coefficient between Alice and Bob. The normalized steering vector is given by
\begin{align}
\mathbf{h}(\theta)=
\frac{1}{\sqrt{N}}\left[e^{j2\pi\Psi_\theta(1)},...,e^{j2\pi\Psi_\theta(n)},...,e^{j2\pi\Psi_\theta(N)}\right]^T,
\end{align}
where
\begin{align}
\Psi_\theta(n)=-\frac{\left(n-(N+1)/2\right)d \cos\theta}{\lambda}, n=1,\cdots,N,
\end{align}
$\theta$ is the direction angle of arrival or departure, $d$ denotes the antenna spacing, $n$ represents the index of antenna, and $\lambda$ is the wavelength.

Similarly, the received signal at Eve is given by
\begin{align}\label{ye}
y_{Ei}&=\mathbf{u}^H_{Ei}\left[(\sqrt{g_{AIE}}\mathbf{H}^{H}_{IE}\boldsymbol{\Theta}\mathbf{H}_{AI}+\sqrt{g_{AE}}\mathbf{H}^{H}_{AE}) \mathbf{s}+\mathbf{n}_{E}\right]\nonumber\\
&=\mathbf{u}^H_{Ei}\big[\sqrt{\beta_1 P_s}(\sqrt{g_{AIE}}\mathbf{H}^{H}_{IE}\boldsymbol{\Theta}\mathbf{H}_{AI}+\sqrt{g_{AE}}\mathbf{H}^{H}_{AE} )
\mathbf{v}_{1}x_{1} \nonumber\\
&~~~+\sqrt{\beta_2P_s}(\sqrt{g_{AIE}}\mathbf{H}^{H}_{IE}\boldsymbol{\Theta}\mathbf{H}_{AI}+\sqrt{g_{AE}}\mathbf{H}^{H}_{AE})
\mathbf{v}_{2}x_{2}\nonumber\\
&~~~ +\sqrt{\beta_3 P_s}(\sqrt{g_{AIE}}\mathbf{H}^{H}_{IE}
\boldsymbol{\Theta}\mathbf{H}_{AI}+\sqrt{g_{AE}}\mathbf{H}^{H}_{AE})\mathbf{P}_{AN}\mathbf{z}\nonumber\\
&~~~+\mathbf{n}_{E}\big], ~~~i=1,2,
\end{align}
where $\mathbf{u}_{Ei}\in\mathbb{C}^{N_E\times 1}$ is the receive beamforming vector, $\mathbf{H}^H_{IE}=\mathbf{h}(\theta^r_{IE})\mathbf{h}^H(\theta^t_{IE})\in\mathbb{C}^{N_E\times M}$ denotes the IRS-to-Eve channel, $\mathbf{H}^H_{AE}=\mathbf{h}(\theta^r_{AE})\mathbf{h}^H(\theta^t_{AE})\in\mathbb{C}^{N_E\times N_A}$ represents the Alice-to-Eve channel, and $\mathbf{n}_E\sim\mathcal{C}\mathcal{N}(\mathbf{0},\sigma_E^2\mathbf{I}_{N_E})$ denotes AWGN at Eve. $g_{AIE}=g_{AI} g_{IE}$ represents the equivalent path loss coefficient of Alice-to-IRS channel and IRS-to-Eve channel, and $g_{AE}$ denotes the path loss coefficient of Alice-to-Eve channel. In what follows, we assume that $\sigma_B^2=\sigma_E^2=\sigma^2$.

Assuming that the AN  is only transmitted to Eve for interference, then $\mathbf{P}_{AN}$ should satisfy
\begin{align}\label{panzf}
\mathbf{H}_{AI}\mathbf{P}_{AN}=\mathbf{0}_{M\times N_A},~\mathbf{H}_{AB}^{H}\mathbf{P}_{AN}=\mathbf{0}_{N_B\times N_A}.
\end{align}
Let us define a large virtual CM channel as follows
\begin{align}\label{H-CM}
\mathbf{H}_{CM} = \left[ {\begin{array}{*{20}{c}}
\mathbf{H}_{AI}\\
\mathbf{H}_{AB}^{H}
\end{array}} \right],
\end{align}
then $\mathbf{P}_{AN}$ can be casted as
\begin{align}\label{PAN}
\mathbf{P}_{AN}=\mathbf{I}_{N_A}-\mathbf{H}_{CM}^{H}\left[\mathbf{H}_{CM}\mathbf{H}_{CM}^{H}\right]^{\dagger}\mathbf{H}_{CM}.
\end{align}

\subsection{Problem Formulation}
Since the channels from Alice to IRS, Alice to Bob, and IRS to Bob are LOS channels, we have $\text{rank}(\mathbf{H}_{AI})$=1 and  $\text{rank}(\mathbf{H}_{AB})$=1. This means that $\text{rank}(\mathbf{H}_{CM})$ in (\ref{H-CM}) is equal to or smaller than 2. There are at least $N_A-2$  degrees of freedom for AN projection matrix $\mathbf{P}_{AN}$.

In this case, substituting (\ref{PAN}) back into (\ref{yb}) and (\ref{ye}), we can obtain
\begin{align}\label{yban}
y_{Bi}&=\mathbf{u}^H_{Bi}\big[\sqrt{\beta_1 P_s}(\sqrt{g_{AIB}}\mathbf{H}^{H}_{IB}\boldsymbol{\Theta}\mathbf{H}_{AI}+\sqrt{g_{AB}}\mathbf{H}^{H}_{AB})\mathbf{v}_{1}x_{1} \nonumber\\
&~~~+\sqrt{\beta_2P_s}\left(\sqrt{g_{AIB}}\mathbf{H}^{H}_{IB}\boldsymbol{\Theta}\mathbf{H}_{AI}+\sqrt{g_{AB}}\mathbf{H}^{H}_{AB} \right)\mathbf{v}_{2}x_{2}\nonumber\\
&~~~+\mathbf{n}_{B}\big],~~~~i=1,2,
\end{align}
and
\begin{align}\label{yean}
y_{Ei}=&\mathbf{u}^H_{Ei}\big[\sqrt{\beta_1 P_s}(\sqrt{g_{AIE}}\mathbf{H}^{H}_{IE}\boldsymbol{\Theta}\mathbf{H}_{AI}+\sqrt{g_{AE}}\mathbf{H}^{H}_{AE})
\mathbf{v}_{1}x_{1} \nonumber\\
&+\sqrt{\beta_2 P_s}(\sqrt{g_{AIE}}\mathbf{H}^{H}_{IE}\boldsymbol{\Theta}\mathbf{H}_{AI}+\sqrt{g_{AE}}\mathbf{H}^{H}_{AE})
\mathbf{v}_{2}x_{2} \nonumber\\
&+\sqrt{\beta_3 P_s g_{AE}}\mathbf{H}^{H}_{AE}\mathbf{P}_{AN}\mathbf{z} +\mathbf{n}_{E}\big], ~~~~i=1,2.
\end{align}
Let us define
\begin{align}
\mathbf{H}_{B}= \sqrt{g_{AIB}}\mathbf{H}^{H}_{IB}\boldsymbol{\Theta}\mathbf{H}_{AI}+\sqrt{g_{AB}}\mathbf{H}^{H}_{AB},
\end{align}
and
\begin{equation}
\left(
  \begin{array}{ccc}
    \sqrt{\beta_1P_s}\mathbf{u}^H_{B1}\mathbf{H}_{B}\mathbf{v}_1 & \sqrt{\beta_2P_s}\mathbf{u}^H_{B1}\mathbf{H}_{B}\mathbf{v}_2\\
    \sqrt{\beta_1P_s}\mathbf{u}^H_{B2}\mathbf{H}_{B}\mathbf{v}_1 & \sqrt{\beta_2P_s}\mathbf{u}^H_{B2}\mathbf{H}_{B}\mathbf{v}_2\\
  \end{array}
\right)
=\left(
  \begin{array}{ccc}
    A_b & B_b\\
    C_b & D_b\\
  \end{array}
\right).
\end{equation}
Then the received signal in (\ref{yban}) can be rewritten as
\begin{align}
y_B=\left(
  \begin{array}{ccc}
    A_b & B_b\\
    C_b & D_b\\
  \end{array}
\right)
\left(
  \begin{array}{ccc}
    x_1\\
    x_2
  \end{array}
\right) +
\left(
  \begin{array}{ccc}
    \mathbf{u}^H_{B1}\\
    \mathbf{u}^H_{B2}
  \end{array}
\right)\mathbf{n}_{B}.
\end{align}
Then, the achievable rate at Bob is given by
\begin{align}
R_B=&\text{log}_2\text{det}
\Bigg\{\mathbf{I}_2+\left(
  \begin{array}{ccc}
    A_b & B_b\\
    C_b & D_b\\
  \end{array}
\right)\left(
  \begin{array}{ccc}
    A_b & B_b\\
    C_b & D_b\\
  \end{array}
\right)^H\bullet\nonumber\\
&\left[\sigma^2\bullet\left(
  \begin{array}{ccc}
    \mathbf{u}^H_{B1}\\
    \mathbf{u}^H_{B2}
  \end{array}
\right)
\left( \mathbf{u}_{B1}  ~~  \mathbf{u}_{B2}\right)\right]^{-1}\Bigg\}.
\end{align}

Similarly, we define
\begin{align}
\mathbf{H}_{E}= \sqrt{g_{AIE}}\mathbf{H}^{H}_{IE}\boldsymbol{\Theta}\mathbf{H}_{AI}+\sqrt{g_{AE}}\mathbf{H}^{H}_{AE},
\end{align}
and
\begin{equation}
\left(
  \begin{array}{ccc}
    \sqrt{\beta_1P_s}\mathbf{u}^H_{E1}\mathbf{H}_{E}\mathbf{v}_1 & \sqrt{\beta_2P_s}\mathbf{u}^H_{E1}\mathbf{H}_{E}\mathbf{v}_2\\
    \sqrt{\beta_1P_s}\mathbf{u}^H_{E2}\mathbf{H}_{E}\mathbf{v}_1 & \sqrt{\beta_2P_s}\mathbf{u}^H_{E2}\mathbf{H}_{E}\mathbf{v}_2\\
  \end{array}
\right)
=\left(
  \begin{array}{ccc}
    A_e & B_e\\
    C_e & D_e\\
  \end{array}
\right).
\end{equation}
Then (\ref{yean}) can be recasted as
\begin{align}
y_E=&\left(
  \begin{array}{ccc}
    A_e & B_e\\
    C_e & D_e\\
  \end{array}
\right)
\left(
  \begin{array}{ccc}
    x_1\\
    x_2
  \end{array}
\right) +
\left(
  \begin{array}{ccc}
    \mathbf{u}^H_{E1}\\
    \mathbf{u}^H_{E2}
  \end{array}
\right)\sqrt{\beta_3 P_s g_{AE}}\bullet\nonumber\\
&\mathbf{H}^{H}_{AE}\mathbf{P}_{AN}\mathbf{z}+\left(
  \begin{array}{ccc}
    \mathbf{u}^H_{E1}\\
    \mathbf{u}^H_{E2}
  \end{array}
\right)\mathbf{n}_{E}.
\end{align}
The achievable rate at Eve is
\begin{align}
R_E=&\text{log}_2\text{det}
\Bigg\{\mathbf{I}_2+\left(
  \begin{array}{ccc}
    A_e & B_e\\
    C_e & D_e\\
  \end{array}
\right)\left(
  \begin{array}{ccc}
    A_e & B_e\\
    C_e & D_e\\
  \end{array}
\right)^H\bullet\nonumber\\
&\Bigg[\left(
  \begin{array}{ccc}
    \mathbf{u}^H_{E1}\\
    \mathbf{u}^H_{E2}
  \end{array}
\right)
\beta_3 P_s g_{AE}\mathbf{H}^{H}_{AE}\mathbf{P}_{AN}\mathbf{P}^H_{AN}\mathbf{H}_{AE}
\left(\mathbf{u}_{E1}~~\mathbf{u}_{E2}\right) \nonumber\\
&+\sigma^2\bullet\left(
  \begin{array}{ccc}
    \mathbf{u}^H_{E1}\\
    \mathbf{u}^H_{E2}
  \end{array}
\right)
\left(\mathbf{u}_{E1}~~\mathbf{u}_{E2}\right)\Bigg]^{-1}\Bigg\}.
\end{align}

The achievable SR $R_s$ can be given by
\begin{align}
R_s=\max\left\{0, R_B-R_E\right\}.
\end{align}
Then, the SR optimization problem is given as follows
\begin{subequations}\label{SR}
\begin{align}
&\max_{\mathbf{u}_{B1},\mathbf{u}_{B2},\boldsymbol{\Theta}}~~~
R_s (\mathbf{u}_{B1},\mathbf{u}_{B2},\boldsymbol{\Theta})\\
&~~~~~\text{s.t.}~~\mathbf{u}^H_{B1}\mathbf{u}_{B1}=1,~\mathbf{u}^H_{B2}\mathbf{u}_{B2}=1,\\
&~~~~~~~~~~|\Theta_i|=1, ~i=1,\cdots, M.
\end{align}
\end{subequations}

For simplification of the objective function and computational convenience, we convert the SR optimization problem in (\ref{SR}) to the optimization problem of Max-RPS at Bob as follows
\begin{subequations}\label{max_RPS}
\begin{align}
&\max_{\mathbf{u}_{B1},\mathbf{u}_{B2},\boldsymbol{\Theta}}~~~\beta_1 P_{s}\mathbf{u}^H_{B1}\mathbf{H}_{B}\mathbf{v}_{1}\mathbf{v}_{1}^{H}\mathbf{H}_{B}^{H}\mathbf{u}_{B1}\nonumber\\
&~~~~~~~~~~~~+\beta_{2}P_{s}\mathbf{u}^H_{B2}\mathbf{H}_{B}\mathbf{v}_{2}\mathbf{v}_{2}^{H}
\mathbf{H}_{B}^{H}\mathbf{u}_{B2}\\
&~~~~~\text{s.t.}~~\mathbf{u}^H_{B1}\mathbf{u}_{B1}=1,~\mathbf{u}^H_{B2}\mathbf{u}_{B2}=1,\\
&~~~~~~~~~~|\Theta_i|=1,~i=1,\cdots, M.
\end{align}
\end{subequations}
Solving this problem is a challenge since the unit modulus constraint is difficult to handle. In this case, two optimal alternating methods are proposed to design the receive beamforming vectors and IRS PSM.
\section{Proposed Max-RPS-GAO scheme}\label{S3}
In this section, the transmit beamforming vectors are designed firstly. Then, we will propose a GAO-based Max-RPS method to obtain the confidential message RBF vectors $\mathbf{u}_{B1}$, $\mathbf{u}_{B2}$ and IRS PSM $\boldsymbol{\Theta}$ by alternately optimizing one and fixing another.
\subsection{Design of the transmit beamforming vectors}
Firstly, to fix CM precoding vectors $\mathbf{v}_1$ and $\mathbf{v}_2$, channel matrix $\mathbf{H}_{CM}$ in (\ref{H-CM}) is first decomposed as the singular-value decomposition (SVD)
\begin{align}
\mathbf{H}_{CM} = \widetilde{\mathbf{U}}\mathbf{\Sigma}_{CM}\widetilde{\mathbf{V}}^H
&=\sum_{i=1}^2\widetilde{\mathbf{u}}_i\widetilde{\mathbf{\lambda}}_i\widetilde{\mathbf{v}}_i^H,
\end{align}
where $\widetilde{\mathbf{U}}$ and $\widetilde{\mathbf{V}}$ are unitary matrices, and $\mathbf{\Sigma}_{CM}$ is a matrix containing the singular values of $\mathbf{H}_{CM}$ and along its main diagonal.
Let us define the transmit beamforming vector $\mathbf{v}_1=\widetilde{\mathbf{v}}_1$, $\mathbf{v}_2=\widetilde{\mathbf{v}}_2$, where $\widetilde{\mathbf{v}}_1$ and $\widetilde{\mathbf{v}}_2$ can be obtained from the eigenvectors corresponding to the first two largest eigenvalues in $\mathbf{\Sigma}_{CM}$, respectively.


\subsection{Optimize RBF vectors $\mathbf{u}_{B1}$ and $\mathbf{u}_{B2}$ given IRS PSM $\boldsymbol{\Theta}$}
Let us define a large virtual receive channel as follows
\begin{align}\label{H-BR}
\mathbf{H}_{BR} = \left[ {\begin{array}{*{20}{c}}
\mathbf{H}_{IB}^H~~
\mathbf{H}_{AB}^{H}
\end{array}} \right].
\end{align}
To obtain the initial values $\mathbf{u}^{(0)}_{B1}$ and $\mathbf{u}^{(0)}_{B2}$, channel matrix $\mathbf{H}_{BR}$ is first decomposed as the SVD criterion
\begin{align}
\mathbf{H}_{BR} = \widehat{\mathbf{U}}\mathbf{\Sigma}_{BR}\widehat{\mathbf{V}}^H
&=\sum_{i=1}^2\widehat{\mathbf{u}}_i\widehat{\mathbf{\lambda}}_i\widehat{\mathbf{v}}_i^H,
\end{align}
where $\widehat{\mathbf{U}}$ and $\widehat{\mathbf{V}}$ are unitary matrices, and $\mathbf{\Sigma}_{BR}$ represents a matrix containing the singular values of $\mathbf{H}_{BR}$ and along its main diagonal. We define the RBF $\mathbf{u}^{(0)}_{Bi}=\widehat{\mathbf{u}}_{i}$, where $\widehat{\mathbf{u}}_{i}$ can be derived from the eigenvectors corresponding to the first two largest eigenvalues in $\mathbf{\Sigma}_{BR}$.

To simplify the expression of RPS related to the receive beamforming vectors, we regard $\boldsymbol{\Theta}$ as a given constant matrix, and the optimization problem of Max-RPS at Bob related to RBF $\mathbf{u}_{B1}$ can be simplified to
\begin{subequations}
\begin{align}\label{max_RPS_ub1}
&\max_{\mathbf{u}_{B1}}~~\beta_1 P_{s}\mathbf{u}^H_{B1}\mathbf{H}_{B}\mathbf{v}_{1}\mathbf{v}_{1}^{H}\mathbf{H}_{B}^{H}\mathbf{u}_{B1}\\
&~~\text{s.t.}~~~~ \mathbf{u}^H_{B1}\mathbf{u}_{B1}=1.
\end{align}
\end{subequations}
According to the Rayleigh-Ritz theorem \cite{Golub1996Matrix}, the optimal $\mathbf{u}_{B1}$ can be obtained from the eigenvector corresponding to the largest eigenvalue of the matrix $\beta_1 P_{s}\mathbf{H}_{B}\mathbf{v}_{1}\mathbf{v}_{1}^{H}\mathbf{H}_{B}^{H}$.

Similarly, given the determined or known $\mathbf{u}_{B1}$ and $\boldsymbol{\Theta}$, the subproblem to optimize $\mathbf{u}_{B2}$ can be expressed as follows:
\begin{subequations}
\begin{align}\label{max_RPS_ub2}
&\max_{\mathbf{u}_{B2}}~~\beta_2 P_{s}\mathbf{u}^H_{B2}\mathbf{H}_{B}\mathbf{v}_{2}\mathbf{v}_{2}^{H}\mathbf{H}_{B}^{H}\mathbf{u}_{B2}\\
&~~\text{s.t.}~~~~ \mathbf{u}^H_{B2}\mathbf{u}_{B2}=1.
\end{align}
\end{subequations}
In accordance with the Rayleigh-Ritz theorem, the optimal $\mathbf{u}_{B2}$ can be obtained from the eigenvector corresponding to the largest eigenvalue of the matrix $\beta_2 P_{s}\mathbf{H}_{B}\mathbf{v}_{2}\mathbf{v}_{2}^{H}\mathbf{H}_{B}^{H}$.


\subsection{Optimize IRS PSM $\boldsymbol{\Theta}$ given the RBF vectors $\mathbf{u}_{B1}$ and $\mathbf{u}_{B2}$}
To simplify the expression of RPS in this subsection, we regard $\mathbf{u}_{B1}$ and $\mathbf{u}_{B2}$ as the given constant vectors and define the IRS phase-shift vector $\boldsymbol{\theta}$ containing all the elements on the diagonal of $\boldsymbol{\Theta}$, i.e.,
\begin{align}\label{thetade}
\boldsymbol{\Theta}=\text{diag}\{\boldsymbol{\theta}\},
\end{align}
where
\begin{align}
\boldsymbol{\theta}=\left[e^{j\varphi_1},\cdots,e^{j\varphi_i},\cdots, e^{j\varphi_M}\right]^T.
\end{align}
Letting $\theta_i=e^{j\varphi_i}$ be the $i$-th element of $\boldsymbol{\theta}$,  the IRS phase-shift vector $\boldsymbol{\theta}$ should satisfy
\begin{align}\label{thetacon}
|\theta_{i}|=1,~\arg(\theta_{i})\in[0,2\pi),~i=1,\cdots, M.
\end{align}

In what follows, let us  define
\begin{align}
&\mathbf{h}^H_{b1}=\mathbf{u}^H_{B1}\mathbf{H}^{H}_{IB},
\mathbf{h}_{A1}=\mathbf{H}_{AI}\mathbf{v}_{1},\\
&\mathbf{h}^H_{b2}=\mathbf{u}^H_{B2}\mathbf{H}^{H}_{IB},
\mathbf{h}_{A2}=\mathbf{H}_{AI}\mathbf{v}_{2},\\
&t_{h_{AIB1}}(\boldsymbol{\theta})=\sqrt{\beta_1 P_s g_{AIB}}\mathbf{h}_{b1}^{H}\text{diag}\{\boldsymbol{\theta}\}\mathbf{h}_{A1}\nonumber\\
&~~~~~~~~~~~\overset{(a)}{=}\underbrace{\sqrt{\beta_1 P_s g_{AIB}}\mathbf{h}_{b1}^{H}\text{diag}\{\mathbf{h}_{A1}\}}_{\mathbf{w}^H_{h_{AIB1}}}\boldsymbol{\theta},\\
&t_{h_{AIB2}}(\boldsymbol{\theta})=\sqrt{\beta_2 P_s g_{AIB}}\mathbf{h}_{b2}^{H}\text{diag}\{\boldsymbol{\theta}\}\mathbf{h}_{A2}\nonumber\\
&~~~~~~~~~~~\overset{(a)}{=}\underbrace{\sqrt{\beta_2 P_s g_{AIB}}\mathbf{h}_{b2}^{H}\text{diag}\{\mathbf{h}_{A2}\}}_{\mathbf{w}^H_{h_{AIB2}}}\boldsymbol{\theta},\\
&t_{h_{AB1}}=\sqrt{\beta_1 P_s g_{AB}}\mathbf{u}_{B1}^{H}\mathbf{H}_{AB}^{H}\mathbf{v}_1,\\
&t_{h_{AB2}}=\sqrt{\beta_2 P_s g_{AB}}\mathbf{u}_{B2}^{H}\mathbf{H}_{AB}^{H}\mathbf{v}_2,
\end{align}
where $(a)$ holds due to the fact that $\text{diag}\{\mathbf{a}\}\mathbf{b}=\text{diag}\{\mathbf{b}\}\mathbf{a}$.
Then, the optimization problem of Max-RPS at Bob related to $\boldsymbol{\theta}$ can be expressed as follows
\begin{subequations}\label{P11}
\begin{align}
&\max_{\boldsymbol{\theta}}~~\left(\mathbf{w}^H_{h_{AIB1}}\boldsymbol{\theta}+t_{h_{AB1}}\right)^H
\left(\mathbf{w}^H_{h_{AIB1}}\boldsymbol{\theta}+t_{h_{AB1}}\right)\nonumber\\
&~~~~~~~+\left(\mathbf{w}^H_{h_{AIB2}}\boldsymbol{\theta}+t_{h_{AB2}}\right)^H\left(\mathbf{w}^H_{h_{AIB2}}
\boldsymbol{\theta}+t_{h_{AB2}}\right)\\
&~~\text{s.t.}~~~(\ref{thetacon}).
\end{align}
\end{subequations}
The objective function in (\ref{P11}) can be rewritten as
\begin{align}
f(\boldsymbol{\theta})&=\boldsymbol{\theta}^H(\mathbf{w}_{h_{AIB1}}\mathbf{w}^H_{h_{AIB1}}+
\mathbf{w}_{h_{AIB2}}\mathbf{w}^H_{h_{AIB2}}  )\boldsymbol{\theta}\nonumber\\
&+(t_{h_{AB1}}\mathbf{w}^H_{h_{AIB1}}+t_{h_{AB2}}\mathbf{w}^H_{h_{AIB2}})\boldsymbol{\theta}+
\boldsymbol{\theta}^H(\mathbf{w}_{h_{AIB1}}t_{h_{AB1}}\nonumber\\
&+\mathbf{w}_{h_{AIB2}}t_{h_{AB2}})+t^H_{h_{AB1}}t_{h_{AB1}}+t^H_{h_{AB2}}t_{h_{AB2}}.
\end{align}
To obtain the optimal IRS phase-shift vector, we need to compute the derivative of $f(\boldsymbol{\theta})$ with respect to $\boldsymbol{\theta}$,
\begin{align}
\frac{\partial f(\boldsymbol{\theta})}{\partial\boldsymbol{\theta}}&=\left(\mathbf{w}_{h_{AIB1}}\mathbf{w}^H_{h_{AIB1}}+
\mathbf{w}_{h_{AIB2}}\mathbf{w}^H_{h_{AIB2}}  \right)^T\boldsymbol{\theta}^*\nonumber\\
&+\left(t_{h_{AB1}}\mathbf{w}^H_{h_{AIB1}}+t_{h_{AB2}}\mathbf{w}^H_{h_{AIB2}}\right)^T=0,
\end{align}
which yields
\begin{align}\label{opt_theta}
\boldsymbol{\theta}&=-\left(\mathbf{w}_{h_{AIB1}}\mathbf{w}^H_{h_{AIB1}}+\mathbf{w}_{h_{AIB2}}\mathbf{w}^H_{h_{AIB2}}  \right)^\dag\nonumber\\
&~~~\bullet\left(\mathbf{w}_{h_{AIB1}}t^H_{h_{AB1}}+\mathbf{w}_{h_{AIB2}}t^H_{h_{AB2}}\right).
\end{align}

\subsection{Overall Algorithm}
So far, we have completed the design of RBF vectors and PSM. The iterative idea of the proposed Max-RPS-GAO algorithm is summarized as follows: given a fixed IRS PSM $\boldsymbol{\Theta}$, the corresponding RBF vectors can be computed in a closed-form expression iteratively; given the RBF vectors $\mathbf{u}_{B1}$ and $\mathbf{u}_{B2}$, $\boldsymbol{\theta}$ can be determined by (\ref{opt_theta}) in a closed-form expression; reform $\boldsymbol{\theta}=\exp\{j\angle(\boldsymbol{\theta})\}$, $\boldsymbol{\Theta}=\text{diag}\{\boldsymbol{\theta}\}$. The alternative iteration process among $\mathbf{u}_{B1}$, $\mathbf{u}_{B2}$, and $\boldsymbol{\Theta}$ is repeated until the termination condition is met, i.e., $R_s^{(p)}-R_s^{(p-1)}$ with $p$ being the iteration index.

The computational complexities of proposed Max-RPS-GAO algorithm is
\begin{align}
&\mathcal{O}\Big(D\big[ M^3+(2N_B+5)\label{GAI-Complexity} M^2+(2N_B N_A +2N_A+\nonumber\\
&~~~~~ 2N_B+2)M+(2N_B^3+2N_B^2+2N_BN_A)\big]\Big)
\end{align}
float-point operations (FLOPs), where $D$ denotes the maximum number of alternating iteration.

\section{Proposed low-complexity Max-RPS-ZF scheme}\label{S4}

In this section, a maximizing RPS alternate optimization method is proposed to reduce computational complexity. In accordance with the zero-forcing principle, the receive beamforming vectors $\mathbf{u}_{B1}$ and $\mathbf{u}_{B2}$  can be determined by
\begin{align}\label{nsp}
&\mathbf{u}^H_{B1}\mathbf{H}_{AB}^{H}=\mathbf{0}_{1\times N_A}, ~\mathbf{u}^H_{E1}\mathbf{H}_{AE}^{H}=\mathbf{0}_{1\times N_A},\\
&\mathbf{u}^H_{B2}\mathbf{H}^H_{IB}=\mathbf{0}_{1\times M}, ~\mathbf{u}^H_{E2}\mathbf{H}^H_{IE}=\mathbf{0}_{1\times M},
\end{align}
which means that the RBF vector $\mathbf{u}_{B1}$ is only used to receive the CM reflected from the IRS, and $\mathbf{u}_{B2}$ is used to receive the CM  through the direct path. The received signals at Bob and Eve are
\begin{align}\label{yban_zf}
y_{B1}&=\mathbf{u}^H_{B1}\big[\sqrt{\beta_1 P_s g_{AIB}}\mathbf{H}^{H}_{IB}\boldsymbol{\Theta}\mathbf{H}_{AI}\mathbf{v}_{1}x_{1} \nonumber\\
&~~~+\sqrt{\beta_2 P_s g_{AIB}}\mathbf{H}^{H}_{IB}\boldsymbol{\Theta}\mathbf{H}_{AI} \mathbf{v}_{2}x_{2} +\mathbf{n}_{B}\big],\\
y_{B2}&=\mathbf{u}^H_{B2}\big[\sqrt{\beta_1 P_s g_{AB}}\mathbf{H}^{H}_{AB}\mathbf{v}_{1}x_{1} \nonumber\\
&~~~+\sqrt{\beta_2 P_s g_{AB}}\mathbf{H}^{H}_{AB}\mathbf{v}_{2}x_{2} +\mathbf{n}_{B}\big],
\end{align}
and
\begin{align}\label{yean_zf}
y_{E1}&=\mathbf{u}^H_{E1}\big[\sqrt{\beta_1 P_s g_{AIE}}\mathbf{H}^{H}_{IE}\boldsymbol{\Theta}\mathbf{H}_{AI}\mathbf{v}_{1}x_{1} \nonumber\\
&~~~+\sqrt{\beta_2 P_s g_{AIE}}\mathbf{H}^{H}_{IE}\boldsymbol{\Theta}\mathbf{H}_{AI}\mathbf{v}_{2}x_{2} +\mathbf{n}_{E}\big], \\
y_{E2}&=\mathbf{u}^H_{E2}\big[\sqrt{\beta_1 P_s g_{AE}}\mathbf{H}^{H}_{AE}\mathbf{v}_{1}x_{1} +\sqrt{\beta_2 P_s g_{AE}} \mathbf{H}^{H}_{AE} \mathbf{v}_{2}x_{2}\nonumber\\
&~~~+\sqrt{\beta_3 P_s g_{AE}}\mathbf{H}^{H}_{AE}\mathbf{P}_{AN}\mathbf{z} +\mathbf{n}_{E}\big],
\end{align}
respectively.

Then, the optimization problem of Max-RPS in (\ref{max_RPS}) can be casted as follows
\begin{subequations}\label{maxRP_zf}
\begin{align}
&\max_{\mathbf{u}_{B1},\mathbf{u}_{B2},\boldsymbol{\Theta}}~~\beta_1 P_{s}\mathbf{u}^H_{B1}\mathbf{H}_{B}\mathbf{v}_{1}\mathbf{v}_{1}^{H}\mathbf{H}_{B}^{H}\mathbf{u}_{B1}\nonumber\\
&~~~~~~~~~~~+\beta_{2}P_{s}\mathbf{u}^H_{B2}\mathbf{H}_{B}\mathbf{v}_{2}\mathbf{v}_{2}^{H}
\mathbf{H}_{B}^{H}\mathbf{u}_{B2}\\
&~~~~~\text{s.t.}~~\mathbf{u}^H_{B1}\mathbf{H}_{AB}^{H}=\mathbf{0}_{1\times N_A},~~\mathbf{u}^H_{B2}\mathbf{H}^H_{IB}=\mathbf{0}_{1\times M},\\
&~~~~~~~~~~~\mathbf{u}^H_{B1}\mathbf{u}_{B1}=1,~\mathbf{u}^H_{B2}\mathbf{u}_{B2}=1,\\
&~~~~~~~~~~~|\Theta_i|=1,~i=1,\cdots, M.
\end{align}
\end{subequations}
In what follows, we consider to optimal RBF vectors and IRS PSM by alternately calculating $\mathbf{u}_{B1}$, $\mathbf{u}_{B2}$, and $\boldsymbol{\Theta}$.
\subsection{Optimize RBF vectors $\mathbf{u}_{B1}$ and $\mathbf{u}_{B2}$  given IRS PSM $\boldsymbol{\Theta}$}
In this section, we regard $\boldsymbol{\Theta}$ as a given constant matrix, and the optimization problem of Max-RPS at Bob related to RBF $\mathbf{u}_{B1}$ can be simplified to
\begin{subequations}
\begin{align}
&\max_{\mathbf{u}_{B1}}~\beta_1 P_{s}g_{AIB}\mathbf{u}^H_{B1}\mathbf{H}^{H}_{IB}\boldsymbol{\Theta}\mathbf{H}_{AI}\mathbf{v}_1\mathbf{v}^H_1\mathbf{H}^H_{AI}
\boldsymbol{\Theta}^H\mathbf{H}_{IB}\mathbf{u}_{B1}\\
&~~\text{s.t.}~~~ \mathbf{u}^H_{B1}\mathbf{u}_{B1}=1.
\end{align}
\end{subequations}
Based on the Rayleigh-Ritz theorem, the optimal RBF vector $\mathbf{u}_{B1}$ can be derived from the eigenvector corresponding to the largest eigenvalue of the matrix $\beta_1 P_{s}g_{AIB}\mathbf{H}^{H}_{IB}\boldsymbol{\Theta}\mathbf{H}_{AI}\mathbf{v}_1\mathbf{v}^H_1\mathbf{H}^H_{AI}
\boldsymbol{\Theta}^H\mathbf{H}_{IB}$.

Similarly, given the determined or known $\mathbf{u}_{B1}$ and $\boldsymbol{\Theta}$, the subproblem to optimize $\mathbf{u}_{B2}$ can be expressed as follows
\begin{subequations}
\begin{align}
&\max_{\mathbf{u}_{B2}}~\beta_{2}P_{s}g_{AB}\mathbf{u}^H_{B2}\mathbf{H}^H_{AB}\mathbf{v}_{2}
\mathbf{v}_{2}^{H}\mathbf{H}_{AB}\mathbf{u}_{B2}\\
&~~\text{s.t.}~~~ \mathbf{u}^H_{B2}\mathbf{u}_{B2}=1.
\end{align}
\end{subequations}
In accordance with the Rayleigh-Ritz theorem, the optimal $\mathbf{u}_{B2}$ can be obtained from the eigenvector corresponding to the largest eigenvalue of the matrix $\beta_{2}P_{s}g_{AB}\mathbf{H}^H_{AB}\mathbf{v}_{2}\mathbf{v}_{2}^{H}\mathbf{H}_{AB}$.

\subsection{Optimize IRS PSM $\boldsymbol{\Theta}$ given the RBF vectors $\mathbf{u}_{B1}$ and $\mathbf{u}_{B2}$}

Since the second item of the objective function in (\ref{maxRP_zf}) are independent of $\boldsymbol{\theta}$, the subproblem to optimize $\boldsymbol{\theta}$ can be expressed as follows
\begin{subequations}
\begin{align}
&\max_{\boldsymbol{\theta}}~~\boldsymbol{\theta}^H\mathbf{w}_{h_{AIB1}}\mathbf{w}^H_{h_{AIB1}}\boldsymbol{\theta}\\
&~~\text{s.t.}~~~(\ref{thetacon}).
\end{align}
\end{subequations}
According to the Rayleigh-Ritz theorem, the optimal $\boldsymbol{\theta}$ can be derived from the eigenvector corresponding to the largest eigenvalue of the matrix $\mathbf{w}_{h_{AIB1}}\mathbf{w}^H_{h_{AIB1}}$.

\subsection{Overall Algorithm}
First, we can obtain a new objective function according to the zero-forcing criterion. Then, fix IRS PSM $\boldsymbol{\Theta}$ and use the Rayleigh-Ritz theorem to obtain the  RBF vectors $\mathbf{u}_{B1}$, $\mathbf{u}_{B2}$. In the following, fix the $\mathbf{u}_{B1}$ and $\mathbf{u}_{B2}$, convert the objective variable $\boldsymbol{\Theta}$ into the phase shift vector $\boldsymbol{\theta}$, and obtain the optimal $\boldsymbol{\theta}$ by Rayleigh-Ritz theorem. Since the IRS phase-shift restriction of (\ref{thetacon}), we reform $\boldsymbol{\theta}=\exp\{j\angle(\boldsymbol{\theta})\}$, $\boldsymbol{\Theta}=\text{diag}\{\boldsymbol{\theta}\}$. Finally, loop the above steps, and solve $\mathbf{u}_{B1}$, $\mathbf{u}_{B2}$ and $\boldsymbol{\Theta}$ alternately until the termination condition is satisfied.

The computational complexity of the proposed Max-RPS-ZF algorithm is
\begin{align}
\mathcal{O}\Big(L[ M^3+(2N_B+1)\label{NSP-Complexity} M^2+(2N_B^3+2N_B^2)]\Big)
\end{align}
FLOPs, where $L$ denotes the maximum number of alternating iteration.

\section{Simulation Results}\label{S5}
In this section, simulations are presented to evaluate the performance of the proposed two schemes. System default parameters are set as follows: $P_s=30$ dBm, $\beta_1 = \beta_2 = 0.4,~ \beta_3 = 0.2$, $\sigma_B^2=\sigma_E^2$,
 $N_A=16$, $N_B=N_E=4$, $M = 80$. The distances of Alice-to-IRS, Alice-to-Bob, and Alice-to-Eve are set as $d_{AI}=10$ m, $d_{AB}=50$ m, and $d_{AE}=50$ m, respectively. The angles of departure (AoDs) of each channel are set as $\theta^{t}_{AI}=5\pi/36$, $\theta^{t}_{AB}=11\pi/36$, and $\theta^{t}_{AE}=\pi/3$, respectively. Since the AoD and distance of each channel are given, the channel state information (CSI) of each channel in IRS-aided DM system can be determined.

In what follows, there are two benchmark schemes used to compare with our proposed methods:
\begin{enumerate}
\item \textbf{Random Phase}: The phase for each reflection element of IRS is uniformly and independently generated from $[0,2\pi)$.
\item \textbf{No-IRS}: We assume that the IRS related channel matrices are zero matrices, i.e., $\textbf{H}_{AI}=\textbf{0}$, $\textbf{H}_{IB}=\textbf{0}$, and $\textbf{H}_{IE}=\textbf{0}$.
\end{enumerate}

\begin{figure}[htbp]
  \centering
  \includegraphics[width=0.52\textwidth]{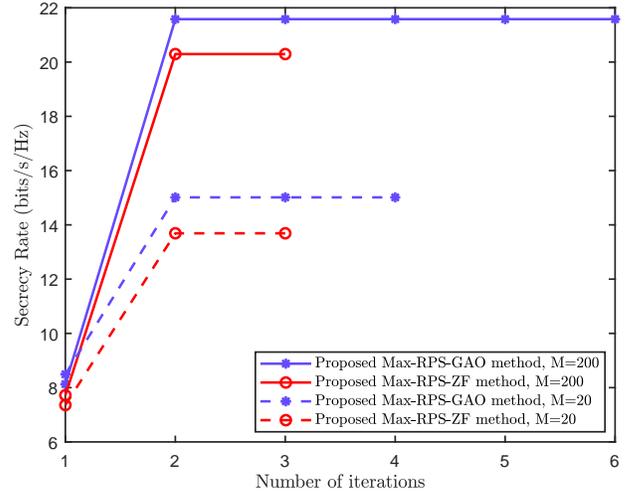}
  \caption{Convergence of proposed algorithms at different number of IRS phase-shift elements. }\label{SR_CON}
\end{figure}

Fig.~\ref{SR_CON} plots the curves of SR versus number of iterations for different number of phase shifters $M=20, 200$. It is observed from Fig.~\ref{SR_CON} that as the number of iterations increases, the SR performances of the proposed Max-RPS-GAO and Max-RPS-ZF algorithms increase gradually and finally converge to a SR floor. In addition, compared with the proposed Max-RPS-GAO algorithm, the convergence rate of proposed Max-RPS-ZF algorithm is faster. From the perspective of computational complexity, when $M=200$, the maximum number of alternating iterations of the proposed Max-RPS-GAO and Max-RPS-ZF algorithms are $D=6$ and $L = 3$, respectively.  According to (\ref{GAI-Complexity}) and (\ref{NSP-Complexity}), when the number of $M$ tends to large scale, the computational complexity of the proposed Max-RPS-ZF algorithm is much lower than that of the proposed Max-RPS-GAO algorithm.

\begin{figure}[htbp]
  \centering
  \includegraphics[width=0.52\textwidth]{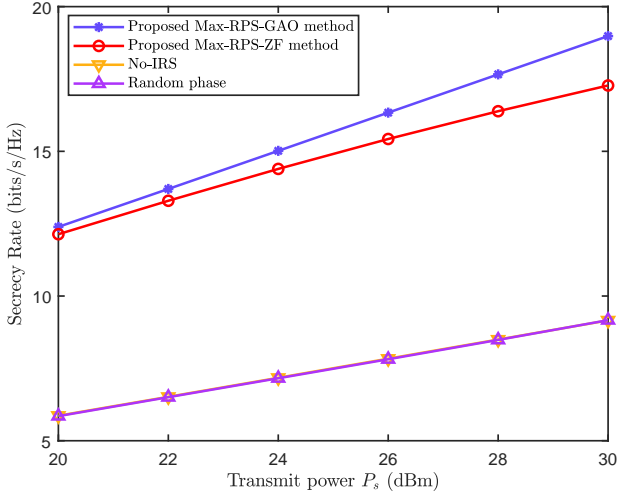}
  \caption{Secrecy rate versus the transmit power $P_s$. }\label{SR_P}
\end{figure}

Fig.~\ref{SR_P} demonstrates the curves of SR versus transmit power $P_s$ of the two proposed methods and two benchmark schemes. It can be seen from this figure that the SR of four schemes increases gradually with the increases of transmit power, and the SR performances of the proposed Max-RPS-GAO and Max-RPS-ZF methods are approximately double that of the no-IRS and random phase schemes regardless of the transmit power. Moreover, the difference of the SR between the no-IRS scheme and the random phase scheme is negligible. This implies that optimizing the IRS phase shift can bring a significant performance improvement.

\begin{figure}[htbp]
  \centering
  \includegraphics[width=0.52\textwidth]{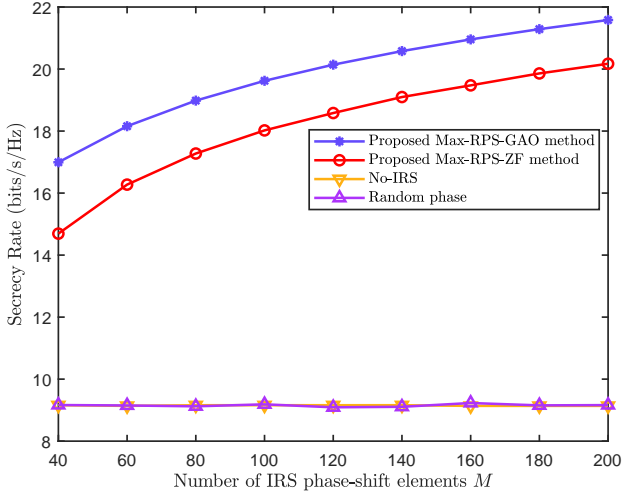}
  \caption{Secrecy rate versus the number of IRS phase-shift elements $M$. }\label{SR_M}
\end{figure}

Fig.~\ref{SR_M} illustrates the curves of SR versus the number of IRS phase shifters elements $M$  of two proposed methods and two benchmark schemes. Compared to the no-IRS and no-PSM optimization schemes, the proposed Max-RPS-GAO and Max-RPS-ZF algorithms can significantly improve the SR performance of the DM system as the number of IRS phase-shift elements $M$ increases. Even with a value of $M$ as $M=40$, the SRs of the Max-RPS-GAO and Max-RPS-ZF methods are increased by about 86\% and 61\%, respectively. Furthermore, it can reflect the superiority of designing and optimizing the PSM of IRS, and the importance of constructing IRS-assisted multipath transmission system.

\begin{figure}[htbp]
  \centering
  \includegraphics[width=0.52\textwidth]{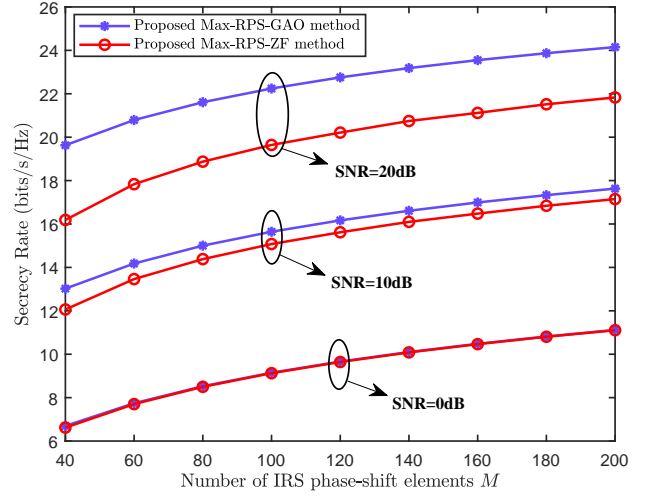}
  \caption{Secrecy rate versus the number of IRS phase-shift elements $M$ in three different SNRs. }\label{SR_SNR}
\end{figure}

Fig.~\ref{SR_SNR} shows the SR versus the number of IRS phase shifters elements $M$ ranging from 40 to 200 in three different SNR scenarios: (1) SNR=0dB, (2) SNR=10dB, and (3) SNR=20dB. It can be seen from the figure that in low SNR region, the difference in SR performance achieved between the Max-RPS-GAO and Max-RPS-ZF algorithms is trivial. However, the difference in SR performance between the two proposed algorithms gradually increases with the increase in SNR, and the difference of SR is about 2 bits/s/Hz when the SNR is equal to 20dB.

\begin{figure}[htbp]
  \centering
  \includegraphics[width=0.52\textwidth]{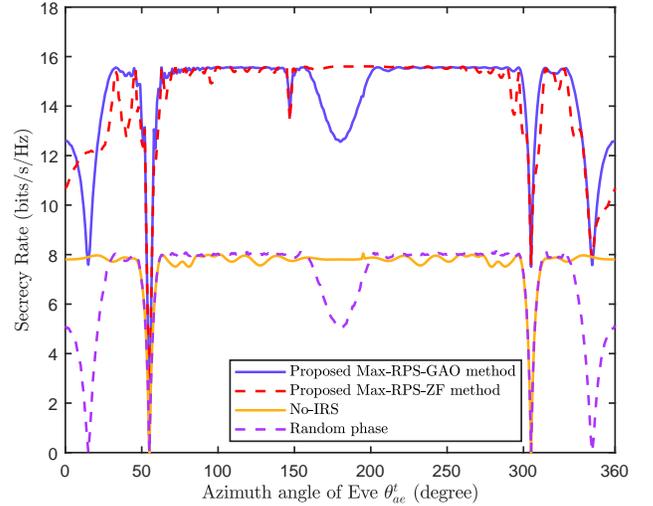}
  \caption{Secrecy rate versus the azimuth angle of Eve $\theta^t_{AE}$. }\label{SR_theta_ae}
\end{figure}

Fig.~\ref{SR_theta_ae} shows the SR versus the azimuth angle $\theta^t_{AE}$ of Eve,  where $\theta^t_{AE}$ changes from 0 to $2\pi$, $\theta^t_{AI}=\pi/12$, and $d_{AB}=d_{AE}=100$m. Since the transmitter and receiver are both linear arrays, the SR performance of $\theta^t_{AE} \in (\pi, 2\pi)$ and $\theta^t_{AE} \in (0, \pi)$  are almost symmetrical to each other. Observing Fig.~\ref{SR_theta_ae}, once Eve and Bob have the same direction angle, i.e., $\theta^t_{AE}=\theta^t_{AB}=11\pi/36$, the SR performance of the four schemes will all decline sharply. This is because Eve is located on the direct path from the Alice to Bob, enabling Eve to eavesdrop on  CMs to the greatest extent. Nevertheless, when $\theta^t_{AE}=\theta^t_{AB}$, the proposed Max-RPS-GAO method still obtains the best SR performance.

\section{Conclusion}\label{S6}
In this paper, we have proposed two alternating iteration optimization schemes, called Max-RPS-GAO algorithm and Max-RPS-ZF algorithm, designing the RBF vectors and PSM in an IRS-aided DM network. The proposed Max-RPS-ZF method uses the zero-forcing criterion to separate two RBF vectors and IRS PSM, such that the two-path CMs from the direct path and the IRS reflected path are independently recovered. Simulation results showed that, compared with the no-IRS-assisted scheme and the no-PSM optimization scheme, the proposed Max-RPS-GAO and Max-RPS-ZF method can significantly improve the SR performance of the DM system as the number of IRS phase shift elements tends to large scale. Compared to the the Max-RPS-GAO, a faster convergence speed and a lower computational complexity can be achieved by the proposed Max-RPS-ZF method. The proposed methods may be applied to the future wireless networks like unmanned aerial vehicle network, satellite communications, vehicle-to-everything, even sixth generation.

\ifCLASSOPTIONcaptionsoff
  \newpage
\fi
\vspace{-3mm}
\bibliographystyle{IEEEtran}
\bibliography{IEEEfull,cite}

\end{document}